\documentclass[aps,prl,9pt, groupedaddress,letter, twocolumn]{revtex4-1}
\usepackage{geometry}
\usepackage{amsmath}
\usepackage{amssymb}
\usepackage{amsfonts}
\usepackage{graphicx}
\usepackage{xcolor}
\usepackage{float}
\usepackage{float}
\usepackage{siunitx}
\usepackage{hyperref}

\DeclareSIUnit{\micron}{\micro\meter}
\DeclareSIUnit{\vel}{\micro\meter\per\second}
\DeclareSIUnit{\diffunit}{\micro\meter\squared\per\second}
\DeclareSIUnit{\mW}{\milli\watt}
\DeclareSIUnit{\ms}{\milli\second}
\DeclareSIUnit{\vol}{\micro\litre}
\DeclareSIUnit{\nm}{\nano\meter}
\DeclareSIUnit{\kHz}{\kilo\hertz}
\definecolor{myBlue}{rgb}{0.1,0.1,0.7}

\begin{document}

\title{Physical Embodiment Enables Information Processing Beyond Explicit Sensing in Active Matter}

\author{Diptabrata Paul}
\author{Frank Cichos}
\email{cichos@physik.uni-leipzig.de}
\affiliation{Molecular Nanophotonics Group, Peter Debye Institute for Soft Matter Physics, Leipzig University, 04103 Leipzig, Germany}
\author{Nikola Milosevic}
\author{Nico Scherf}
\affiliation{Max Planck Institute for Human Cognitive and Brain Sciences, 04103 Leipzig, Germany\\
Center for Scalable Data Analytics and Artificial Intelligence (ScaDS.AI), Dresden/Leipzig, Germany}

\begin{abstract}
Living microorganisms have evolved dedicated sensory machinery to detect environmental perturbations, processing these signals through biochemical networks to guide behavior. Replicating such capabilities in synthetic active matter remains a fundamental challenge. Here, we demonstrate that synthetic active particles can adapt to hidden hydrodynamic perturbations through physical embodiment alone, without explicit sensing mechanisms. Using reinforcement learning to control self-thermophoretic particles, we show that they learn navigation strategies to counteract unobserved flow fields by exploiting information encoded in their physical dynamics. Remarkably, particles successfully navigate perturbations that are not included in their state inputs, revealing that embodied dynamics can serve as an implicit sensing mechanism. This discovery establishes physical embodiment as a computational resource for information processing in active matter, with implications for autonomous microrobotic systems and bio-inspired computation.
\end{abstract}

\maketitle
\section{Introduction}

Navigation in microscopic environment represent one of the most challenging frontiers in robotics, where conventional sensing and control paradigms break down in the face of extreme physical constraints. Yet, motile microorganisms evolved to exhibit surprising computational strategy while navigating such complex environments. At the microscale, in low Reynolds number regime \cite{Purcell1977}, where viscous forces dominate, microorganisms navigate using a dual strategy: they combine cues from dedicated sensory receptors with information inherently encoded in their physical dynamics -- a principle that challenges conventional approaches to autonomous microrobots design \cite{Purcell1977, pfeifer2007howthebody}.

Examples of these dual strategy include E. coli, which possesses chemoreceptors but cannot determine gradient direction through sensing alone due to Brownian noise; instead, their run-and-tumble trajectories create temporal sampling where swimming physics enables gradient detection \cite{berg1993random, ben_jacob2000bacterial, sourjik2002receptor, Bacteria_Feedback_Dufour, Cremer2019_chemotaxis}. Similarly, Paramecium relies on mechanoreceptors to trigger responses, yet its successful navigation arises from how body shape and ciliary dynamics interact with hydrodynamic forces \cite{mehcanoreceptor_paramecium}. Even organisms equipped with sophisticated sensors exploit information from their physical dynamics, enabling morphological computation--where body mechanics transform environmental stimuli into actionable behaviors without explicit sensory cues \cite{adamatzky_slime_2015, hauser2011morphological, judd2019sensing, rorot2022counting, ben2023morphological}. This biological insight reveals a fundamental principle: organisms exploit their physical structure and dynamic interactions with the environment -- embodied dynamics, to extract information about hidden variables and environmental changes \cite{Gupta2021embodied, varma2022embodied, alicea2023embodied}. This paradigm exemplifies embodied intelligence—a concept where physical structure, dynamics, and environmental interactions serve as computational resource to process information which explicit sensors cannot access or that would require prohibitive sensory complexity \cite{nakajima2015information, zhao2024exploring, alicea2023embodied}.

Inspired by nature, artificial microswimmers—a class of synthetic active matter which can mimic the self-propulsion mechanisms of living microorganisms, represent a promising pathway to explore and realize embodied intelligence at the microscale \cite{smart_microswimmer_review_stark, smart_microswimmer_review, MuinosLandin2021}. Yet such microswimmers have not demonstrated this capability. Most current approaches rely entirely on external control systems \cite{Qian2013, Bregulla2014, Khadka2018, Bechinger2024_MARL} or pre-programmed responses \cite{Khadka2018, Wang2023_vortex}, enabling precise steering—such as light-based control of individual swimmers, but lacking both the sensory sophistication and embodied responsiveness observed in biological systems. In this context, reinforcement learning (RL), a framework where an agent gains experience by interacting with its environment, offers a compelling alternative. While simulations have shown RL agents can navigate complex flows \cite{colabrese2017flow, gustavsson2017finding, Biferale2019} with recent experiments demonstrating learned navigation in controlled settings \cite{MuinosLandin2021, Bechinger2024_MARL, mecanna2024critical}, no experimental system has validated whether microswimmers can infer and respond to hidden environmental perturbations using only the information encoded in their physical dynamics.

To this end, we demonstrate that artificial microswimmers can achieve navigation in flow perturbed environments through embodied intelligence, without any explicit flow sensing. Using online reinforcement learning to control self-thermophoretic microrobots, we show that an agent develops a simple radial policy in an inert environment directing them towards a target position. In contrast, the strong advection in a flow-perturbed environment leads the agent to form counteractive policies that oppose the hidden flow fields. Such policies are achieved by exploiting correlations between their states, chosen actions, and the resulting outcomes, effectively extracting hidden information from its physical dynamics.
Remarkably, relying solely on proprioceptive cues and no explicit flow-field information, the agent learns to navigate flows up to four times its propulsion speed within roughly 50 training episodes.

These results highlight an important paradigm for microrobotics: rather than miniaturizing sensors and processors, we can exploit morphological computation where robot bodies become information processors \cite{Gupta2021embodied, alicea2023embodied, hauser2011morphological, dauchot2023morphological}. This enables autonomous microsystems for environments where sensing is impossible - from medical interventions to environmental monitoring - while demonstrating how machine learning can discover the embodied strategies that evolution has crafted over millions of years.

\section{Results}

\begin{figure*}
\includegraphics{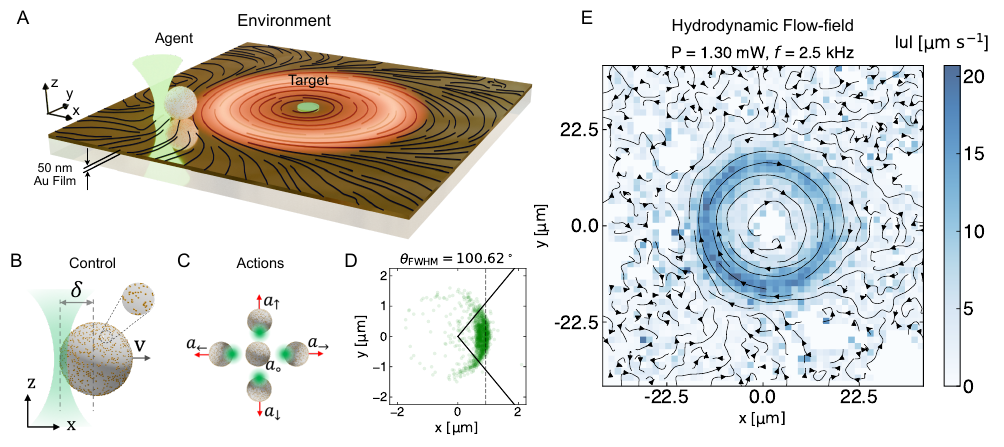}
\caption{\label{f1} \textbf{Experimental realization of reinforcement learning with microswimmer.}
\textbf{A}. The study setup consists of a dilute solution of light activated microswimmer agents in a physically perturbed environment. Perturbations arise from periodic heating of a thin ($\SI{50}{\nm}$) Au film with a focused heating laser, generating hydrodynamic flow fields (black streamlines) that the agents must navigate to reach the target.
\textbf{B}. The control of the microswimmer is enabled by self-thermophoretic motion of an AuNP coated melamine formaldehyde (AuMF) particles ($R = \SI{1.09}{\micron}$) by asymmetric laser illumination ($\delta$ displaced from center).
\textbf{C}. The agent is empowered with five discrete control actions corresponding to heating the particle at specific positions on its circumference (${a_{\uparrow}, a_{\downarrow}, a_{\leftarrow}, a_{\rightarrow}}$) or no heating ($a_{\circ}$).
\textbf{D}. The displacement vector in $x$-$y$ plane for one of the actions for an incident laser power $\text{P}_0 = \SI{0.15}{\mW}$ with a mean displacement of $\SI{0.92}{\micron}$ shown by the vertical dashed black line. The corresponding angular distribution of the displacement vectors is fitted to a normal distribution with $\theta_{\text{FWHM}} = 100.62^{\circ}$ indicated by the solid black lines.
\textbf{E}.  The environmental perturbation is generated due to periodic heating of a thin Au film is characterized by a clockwise hydrodynamic flow field ($\vec{u}$) comprising thermo-osmotic and thermo-viscous effects at frequency $f = \SI{2.5}{\kHz}$ and heating power $\text{P} = \SI{1.30}{\mW}$ traced with $\SI{250}{\nm}$ AuNPs.
}
\end{figure*}

\subsection{Experimental realization}


We developed an experimental platform combining light-activated symmetric microswimmers with real-time reinforcement learning control (Fig. \ref{f1}A). The microswimmers—melamine formaldehyde particles (radius $R = \SI{1.09}{\micron}$) coated with gold nanoparticles—navigate a microfluidic environment while learning to reach target locations despite hydrodynamic perturbations. They are subjected to both inert and physically perturbed environment consisting of hydrodynamic flow fields as indicated by the black streamlines in Fig. \ref{f1}A. The control of the microswimmer motion is achieved by asymmetric illumination with a focused $\lambda = \SI{532}{\nm}$ laser (offset by $\delta \sim R$ from the particle center) inducing self-thermophoretic propulsion away from the laser focus \cite{MuinosLandin2021, Fraenzl2021_ACSNano}(Fig. \ref{f1}B). Such settings is possible via a feedback loop enabled by real-time position detection of the agent in the field of view ($\SI{75}{\micron} \times \SI{75}{\micron}$) coupled to RL program back-end (See Methods and Supplementary Materials 1 for details).


The feedback control with the RL program enables the agent to learn successful navigation strategies by sampling from five discrete actions (see Fig. \ref{f1}C): \begin{equation}
\label{eq1}
\mathcal{A} = \begin{cases} a_{\uparrow}: & v_x = 0,\ v_y = v, \\ a_{\downarrow}:& v_x = 0,\ v_y= -v,\\ a_{\leftarrow}: & v_x = -v,\ v_y= 0, \\ a_{\rightarrow}: & v_x = v,\ v_y= 0,\\ a_{\circ}: & v_x = 0,\ v_y= 0, \end{cases}
\end{equation}
where $v$ denotes the self-propulsion velocity. Beyond deterministic propulsion, the agent's motion experiences stochastic fluctuations from thermal and viscous effects. We quantified this stochasticity by measuring the distribution of displacement vectors ($\Delta\vec{d}$) between successive steps separated by $\tau \approx \SI{160}{\ms}$, as shown in Fig. \ref{f1}D at laser power $\mathrm{P_0} = \SI{0.15}{\mW}$. The distribution reveals a mean displacement of $\approx \SI{0.92}{\micron}$ (vertical dashed line) with angular uncertainty characterized by the opening angle $\theta_{\mathrm{FWHM}} = 2\sqrt{2 \ln 2}\ \sigma_{\theta} = 100.62^{\circ}$ (solid black lines). These values correspond to a mean propulsion speed of $v \approx \SI{5.75}{\vel}$ and Péclet number $\text{Pe} = Rv/D \approx30$, where $D = \SI{0.21}{\diffunit}$ is the diffusion coefficient.


Environmental perturbations are introduced through hydrodynamic flow fields created by periodically heating a thin Au film (\SI{50}{\nm}) with a focused laser beam scanning in a closed circular path. The generated flow field results from two distinct mechanisms: thermo-osmotic attraction toward heated regions and thermo-viscous flow \cite{CichosTOFlow2016, Martin2022, Weinert2008TVFlow, Weinert2008TVFlowPRL}. Thermo-osmotic flows arise from temperature-induced changes in interfacial interaction energy at the heated surface. Thermo-viscous flows are generated by thermal expansion coupled with temperature-dependent fluid viscosity from the scanning-induced thermal wave, creating flow directed opposite to the thermal wave propagation. Scanning the laser in a circular pattern (radius $\SI{15}{\micron}$) at frequency $f = \SI{2.5}{\kHz}$ and power $\text{P} = \SI{1.30}{\mW}$ produces the flow fields visualized in Fig. \ref{f1}E, characterized using $\SI{250}{\nm}$ Au nanoparticles as tracers in dilute aqueous solution (see Supplementary Materials 2 for more details).

For environmental navigation, the agent's state information consists of its position relative to a fixed reference frame and its displacement magnitude, expressed as $s_t=(x_t,\ y_t, |\Delta \vec{d}|)$, where $|\Delta \vec{d}| = d_{t-1} - d_t$ and $d_t$ represents the agent's distance from the target at step $t$. This streamlined state representation captures proprioceptive feedback while avoiding the need for direct sensing of complex or time-varying flow perturbations. Following embodied intelligence principles, the agent exploits its physical interactions with the hydrodynamic environment—as manifested in position updates and displacement measurements—to extract information about hidden environmental features and modify its navigation strategy accordingly. The reward signal that guides policy development depends exclusively on changes in target distance across consecutive time steps:
\begin{equation}
\label{eq2}
r(t) = \begin{cases} -2 & \text{if}\ d_t > d_{t-1}, \\ 1 & \text{if}\ d_t < d_{t-1}, \\ 10 & \text{if}\ d_t \leq d_{\mathrm{th}}. \end{cases}
\end{equation}
where $d_{\text{th}}$ defines the threshold radius around the target. This streamlined, distance-dependent reward framework explicitly captures the navigation goal, promoting target-directed movement without demanding explicit environmental or perturbation data. Additionally, the asymmetric reward design—which penalizes backward motion more severely than it rewards forward progress—limits wasteful exploration and encourages the agent to reduce total navigation duration.

Implementing online RL with this streamlined reward structure in the presence of stochastic embodied dynamics and unobserved flow perturbations demands an algorithm that balances stability with exploratory behavior. We adopt an actor-critic framework that simultaneously optimizes both the policy (actor) and value function (critic). Specifically, we implement proximal policy optimization (PPO), which delivers stable convergence in noisy environments while maintaining computational efficiency—essential characteristics for the demanding experimental conditions we encounter (see Supplementary Materials 3 for more details) \cite{PPOpaper2017}. Similar to motile biological microorganisms, this experimental configuration enables us to examine navigation strategies of an agent functioning under significant sensory limitations and noisy environmental feedback, depending entirely on embodied dynamics to navigate through physically perturbed environments.

\begin{figure*}
\includegraphics[width = 450 pt]{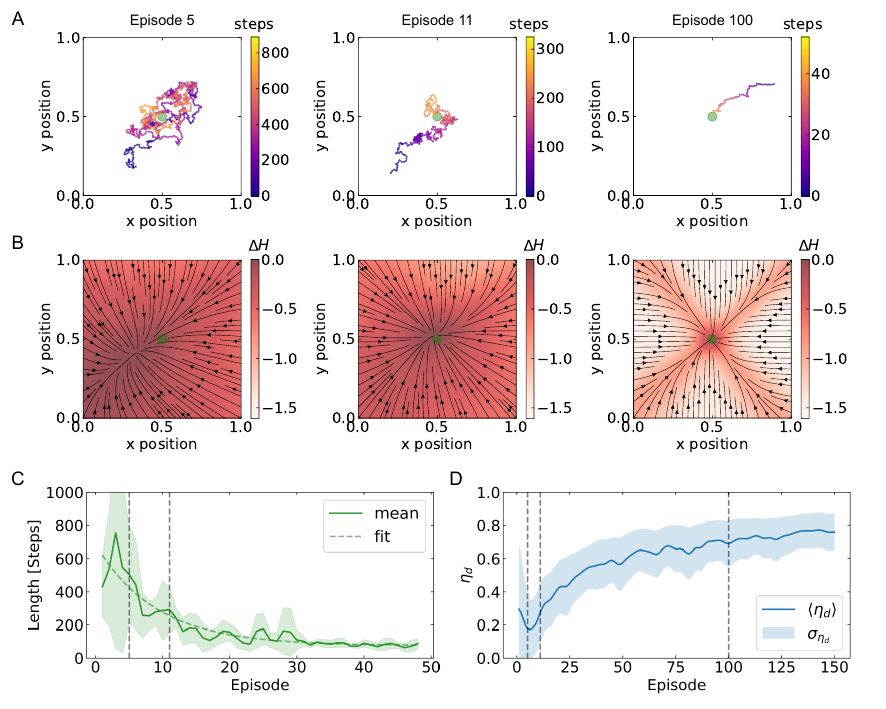}
\caption{\label{f2} \textbf{Learning in an inert environment.}
\textbf{A}. Trajectories of the microswimmer agent at average speed of $v = 6.1\, \mathrm{\mu m/s}$ show that progressive training leads to more deterministic motion towards the target position, indicated by the green circle.
\textbf{B}. The corresponding policy, characterized by Shannon entropy ($\Delta H$) and evaluated relative to the initial state of the agent, indicates evolution to a more deterministic policy. The corresponding expected velocity field $\langle \vec{v}_{\text{A}}\rangle_{\text{inert}}$ evolves towards a radial field, represented by the streamlines.
\textbf{C}. The length of episodes decreases with progressive training episodes and fitted to an exponential decaying function ($\propto \exp(-t/\tau_c)$, $\tau_c = 8.96$ episodes ($\approx 4000$ steps) being the characteristic convergence time. The solid line shows moving mean, the dashed line the corresponding fit and the shaded region indicates the corresponding standard deviation.
\textbf{D}. The path efficiency calculated from the policy increases with progressive training, with the highest value being $\eta_d = 0.76 \pm 0.11$ at the end of the training for 150 episodes. The gray vertical dashed-lines in \textbf{C} and \textbf{D} marks the episode of policy indicated in \textbf{B}.
}
\end{figure*}

\subsection{Learning in an inert environment}

To establish baseline performance, we first evaluate the agent's learning behavior in an unperturbed environment. Each learning episode begins with the agent placed at a random initial position and continues through a sequence of actions until the agent either reaches the target threshold region ($d_t\leq d_{\text{th}}$) or encounters the environment boundary. Fig. \ref{f2}A shows the evolution of navigation trajectories for an agent with propulsion speed $v = \SI{5.75}{\vel}$ ($\text{Pe}\approx 30$) across multiple learning episodes, where gradient colors indicate step progression within each episode.

To quantify policy evolution, we analyze the agent's navigation strategy using Shannon entropy ($\Delta H$) changes relative to the initial state, providing a spatial distribution of decision-making uncertainty:
\begin{equation}
\label{eq3}
\Delta H = -\sum\limits_{a_i \in \mathcal{A}}p_{a_i}\log{p_{a_i}} - H_0.
\end{equation}
The reference entropy $H_0 = \log(0.2)$ represents the initial uniform policy where all actions have equal probability. Fig. \ref{f2}B displays the entropy distribution of the learned policy after corresponding episodes, quantifying the agent's decision-making uncertainty. Progressive training leads to reduced entropy ($\Delta H$) in regions distant from the target, indicating evolution toward more deterministic behavior, while higher entropy persists near the target (green circle), reflecting continued uncertainty in optimal action selection. The anisotropic entropy distribution demonstrates how the discrete action space concentrates uncertainty along diagonal directions, where action probabilities are distributed among competing pairs: $\{\uparrow, \rightarrow\}$, $\{\uparrow, \leftarrow\}$, $\{\downarrow, \rightarrow\}$, and $\{\downarrow, \leftarrow\}$ (detailed analysis in Supplementary Materials 4). These patterns align with PPO's stochastic characteristics \citep{PPOpaper2017}, which approximates entropy-constrained policy updates from TRPO \cite{schulman2015trust}, maintaining exploration through an implicit maximum entropy bias \cite{hu2021actor, muller2024essentially, milosevic2025central}.

To further characterize the learned policy, we evaluate the agent's expected velocity field $\langle \vec{v}_\text{A} \rangle = \sum\limits_{ a_i \in \mathcal{A}}p_{a_i}\vec{v}_i$, where $p_{a_i}$ and $\vec{v}_i$ represent the action probabilities and velocity vectors, respectively. The black streamlines in Fig. \ref{f2}B illustrate the expected velocity field $\langle \vec{v}_\text{A} \rangle_{\text{inert}}$ from policies obtained after corresponding training episodes in the inert environment. This field exhibits a radially inward profile with a sink that progressively aligns with the target position (green circle) (additional details in Supplementary Materials 4).

Episode lengths decrease systematically with training, as demonstrated in Fig. \ref{f2}C, where the solid line represents the mean and the shaded region shows the standard deviation. We extract the characteristic convergence time $\tau_c$ by fitting these decaying values to an exponential function $\propto \exp(-t/\tau_c)$ (green dashed line in Fig. \ref{f2}C), where $t$ denotes episode number. This analysis yields $\tau_c = 8.96$ episodes, equivalent to approximately $4117 \pm 600$ navigational steps (see Supplementary Materials 5 for additional details). We quantify the agent's performance using path efficiency $\eta_d = d_0/\sum_t{\Delta d_t}$, where $d_0$ represents the initial target distance and $\Delta d_t$ is the step length, evaluated in a virtual inert environment. The path efficiency $\eta_d$ increases with training and reaches $\eta_d = 0.76\pm0.11$, as shown in Fig. \ref{f2}D. Consequently, in an inert environment, the learned policy converges to simple, radially inward motion toward the target—requiring no exploitation of embodied dynamics.

\begin{figure*}
\includegraphics[width = 480 pt]{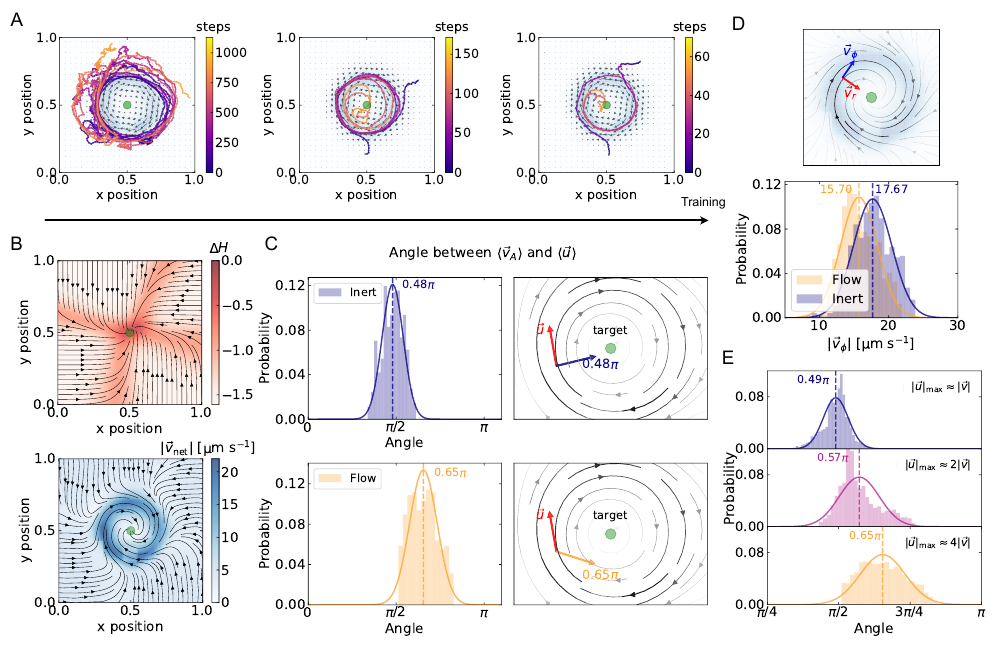}
\caption{\label{f3} \textbf{Learning in flow-perturbed environment.}
\textbf{A}. With progressive training, the agent’s trajectories in the flow-perturbed environment (shows as the background) evolve from circling around the disturbed region to successfully navigation to reach the target.
\textbf{B}. (top) The policy after $60$ training episodes is characterized by the entropy ($\Delta H$) and evaluated with respect to the initial state of the agent. The corresponding expected velocity field $\langle \vec{v}_{\text{A}}\rangle_{\text{flow}}$ is represented by the black streamlines.
(bottom) Net velocity field ($\vec{v}_{\text{net}}$) of the microswimmer agent, computed by adding the $\langle \vec{v}_{\text{A}}\rangle_{\text{flow}}$ with the experimentally measured hydrodynamic flow-field ($\vec{u}$), $\vec{v}_{\text{net}} = \langle \vec{v}_{\text{A}}\rangle_{\text{flow}} + \vec{u}$. The resulting pattern exhibits vortex-like structure centered near the target position (green circle).
\textbf{C}.The policies in the inert and flow-perturbed environments are characterized by the relative angle between $\langle \vec{v}_{\text{A}}\rangle$ and $\vec{u}$. (top) The resulting histogram for the inert environment yields a mean relative angle $0.48 \pi$ for the inert-environment policy, corresponding to a radially inward policy. (bottom) In the flow-perturbed case, the mean angle shifts to $0.65 \pi$, reflecting a counteractive motion against the flow-perturbed region.
\textbf{D}. The net velocity $\vec{v}_{\text{net}}$ in both the inert and flow-perturbed environments is analyzed by decomposing it into tangential ($\vec{v}_{\phi}$) and radial ($\vec{v}_r$) component, with the target located at the origin. The histogram of the $|\vec{v}_{\phi}|$ reveals a  lower magnitude for the flow-perturbed policy ($|\vec{v}_{\phi}|_{\text{mean}}^{\text{flow}}\approx \SI{15.70}{\micron}$) compared to the inert-environment policy ($|\vec{v}_{\phi}|_{\text{mean}}^{\text{inert}}\approx \SI{17.67}{\micron}$). The lower value for the flow-perturbed environment policy is attributed to the counteractive response of trained agent.
\textbf{E}. The flow-perturbed policy quantified by the mean relative angle between $\langle \vec{v}_{\text{A}} \rangle$ and $\vec{u}$ extracted from corresponding histogram increases from $0.49 \pi$ when $|\vec{u}|\approx|\vec{v}|$ to $0.65\pi$ when $|\vec{u}|\approx4|\vec{v}|$, indicating higher counteractive response to stronger perturbation.
}
\end{figure*}

\begin{figure*}
\includegraphics[width = 485 pt]{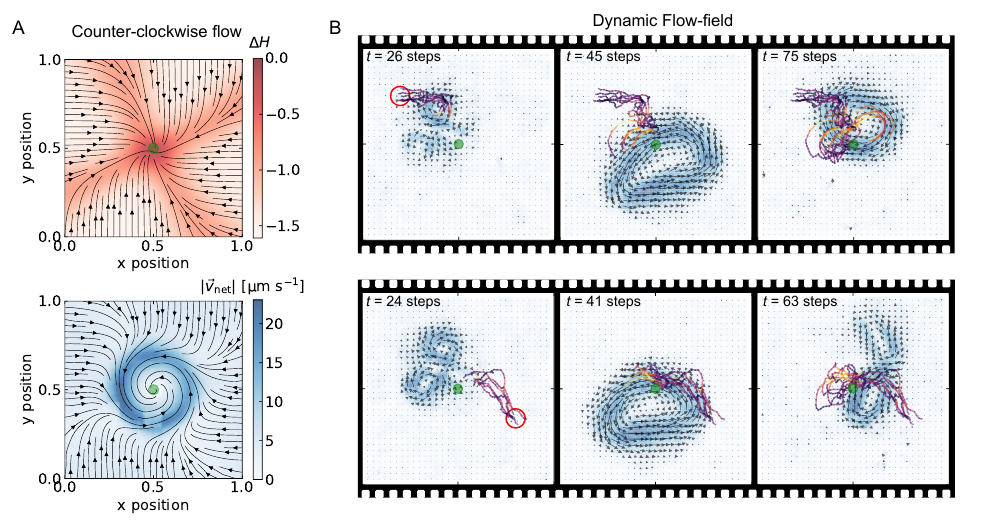}
\caption{\label{f4} \textbf{Learned response to counter-clockwise flow and dynamic flow field.}
\textbf{A}. (top) Policy of an agent trained under the perturbation of counter-clockwise flow-field for $60$ episodes is characterized by the entropy difference ($\Delta H$) from the initial state. The corresponding expected velocity field $\langle \vec{v}_{\text{A}}\rangle_{\text{flow}}$ is represented by the black streamlines.
(bottom) The resulting net velocity profile ($\vec{v}_{\text{net}} = \langle \vec{v}_{\text{A}}\rangle_{\text{flow}} + \vec{u}$), results in a vortex-like pattern covering around the target position (green circle), indicating successful navigation.
\textbf{B}. An agent trained in a dynamic flow-perturbed environment learns an effective policy adapted to time-varying perturbations. Time-series snapshots show trajectories from agents starting at (top) the upper-left and (bottom) the lower-right positions (indicated by the red circle), successfully navigating to the target.
}
\end{figure*}

\subsection{Learning in flow-perturbed environment}

Unlike the inert environment, real-world navigation presents complex challenges where simple radial strategies fail. Environmental perturbations—including boundaries, flow fields, and gradients—define the evolutionary landscape that shapes motile microorganisms' navigation strategies. In such environments, sensory information is typically local, noisy, and incomplete. Under these constraints, successful navigation requires organisms to leverage their physical structure and dynamic interactions with the environment to process information and guide movement.

To examine this principle experimentally, we subject our RL agent to a flow-perturbed environment. The perturbation consists of a clockwise (CW) hydrodynamic flow field generated through periodic heating of the Au film, combining thermo-osmotic and thermo-viscous effects (Fig. \ref{f1}E). As in the inert case, the agent receives only position and displacement information—no direct flow measurements. We tune the agent's propulsion speed ($v$) relative to the maximum flow speed such that $|\vec{v}|/{|\vec{u}|}_{\text{max}}\approx 0.25$, ensuring strong competition between advection and self-propulsion.

The agent's navigation behavior evolves dramatically with training. Initial trajectories exhibit extensive circling around the flow-perturbed region (visible in the background of Fig. \ref{f3}A), reflecting the agent's inability to counteract the strong hydrodynamic forces. Through progressive training, these chaotic paths evolve into increasingly direct routes toward the target (green circle), with the color-coded step count revealing reduced navigation time.

We quantify the learned policy using Shannon entropy ($\Delta H$) relative to the initial uniform distribution, as shown in Fig. \ref{f3}B. The corresponding expected velocity field $\langle\vec{v}_\text{A}\rangle_{\text{flow}}$ (black streamlines) forms a sink coinciding with the target position. Crucially, this policy differs qualitatively from the simple radial pattern $\langle\vec{v}_\text{A}\rangle_{\text{inert}}$ observed in the inert environment (Fig. \ref{f2}B). Instead, the flow-trained policy exhibits a twisted structure—a signature of the agent's learned response to the clockwise perturbation.

To understand this adaptation, we compute the net velocity field $\vec{v}_{\text{net}} = \langle\vec{v}_\text{A}\rangle_{\text{flow}} + \vec{u}$, which combines the agent's learned policy with the external flow. This field exhibits a vortex-like structure centered on the target, accurately representing the trained agent's actual navigation trajectories (Fig. \ref{f3}A).

We quantify the policy's counteractive nature by analyzing the relative angle between the agent's expected velocity $\langle \vec{v}_{\text{A}}\rangle_{\text{flow}}$ and the local flow $\vec{u}$ within the perturbed region. In the inert environment, this relative angle averages $0.48\pi$, consistent with purely radial motion (Fig. \ref{f3}C, top). Under flow perturbation, the mean angle shifts to $0.65\pi$, indicating systematic opposition to the local flow direction (Fig. \ref{f3}C, bottom).

We further analyze this counteractive behavior by decomposing $\vec{v}_{\text{net}}$ into radial ($\vec{v}_{r}$) and tangential ($\vec{v}_{\phi}$) components relative to the target. The effect appears most clearly in the tangential component magnitude $|\vec{v}_{\phi}|$. The flow-trained policy produces a lower mean tangential velocity ($|\vec{v}_{\phi}|_{\text{mean}}^{\text{flow}}\approx \SI{15.70}{\vel}$) compared to the inert policy ($|\vec{v}_{\phi}|_{\text{mean}}^{\text{inert}}\approx \SI{17.67}{\vel}$) (Fig. \ref{f3}D). This reduction reflects the agent's learned strategy to counteract the flow's tangential influence while maintaining progress toward the target.

To test the strength-dependence of this adaptation, we systematically vary the flow field intensity relative to the agent's propulsion speed: $|\vec{u}|_{\text{max}}\approx|\vec{v}|$, $|\vec{u}|_{\text{max}}\approx2|\vec{v}|$, and $|\vec{u}|_{\text{max}}\approx4|\vec{v}|$. The mean relative angle between ${\langle \vec{v}_{\text{A}}\rangle}_{\text{flow}}$ and $\vec{u}$ increases correspondingly from $0.49\pi$ to $0.57\pi$ to $0.65\pi$ (Fig. \ref{f3}E). This progression reveals that pronounced counteractive responses emerge only under strong perturbations, demonstrating the adaptive nature of the learned behavior.

The mechanism underlying this adaptation is fundamentally embodied. Within the perturbed environment, each action's outcome inherently reflects the flow's influence—the agent's observed motion already represents the combined effect of intended propulsion and environmental perturbation (see Supplementary Materials 6). Rather than requiring explicit flow sensing, the agent exploits proprioceptive cues to encode hydrodynamic interactions through correlations between states, chosen actions, and resulting transition probabilities. This correlation constitutes morphological computation, where embodied dynamics serve as a computational resource for inferring and counteracting hidden environmental variables. Effective navigation thus emerges from embodied intelligence, where adaptation to strong perturbations occurs by exploiting the agent's own dynamics to recover information about the hidden flow field.

\subsection{Generalization to reversed and dynamic flows}

The embodied learning framework's generality becomes evident when examining adaptation to varied flow configurations. We first tested counter-clockwise (CCW) flows using identical heating parameters ($f = \SI{2.5}{\kHz}$, $P = \SI{1.30}{\mW}$) but reversed scanning direction, maintaining $|\vec{v}|/{|\vec{u}|}_{\text{max}} \approx 0.25$. The learned policy develops a mirror-image structure—the expected velocity field ${\langle \vec{v}_{\text{A}}\rangle}_{\text{flow}}$ exhibits opposite twist compared to the CW (Fig. \ref{f3}B) case while preserving the sink at the target (Fig. \ref{f4}A (top)). The net velocity profile $\vec{v}_{\text{net}}$ similarly shows a vortical structure around the target (green circle) (\ref{f4}A (bottom)). This systematic reversal confirms that agents discover environment-specific solutions rather than fixed heuristics, validating that embodied dynamics encode sufficient environmental information for adaptive discrimination.


Dynamic flows present a fundamentally different challenge: navigation without predictable correlations between actions and outcomes. We tested this by continuously varying the heating laser's scanning pattern, creating time-varying flows with maximum speeds ${|\vec{u}|}_{\text{max}}\approx \SI{20}{\vel}$—over three times the agent's propulsion capability. Figure \ref{f4}B (top and bottom) demonstrates successful navigation from multiple starting positions (red circles) despite these unpredictable perturbations, shown as the background.

The success in dynamic environments reveals a crucial principle of embodied learning. Lacking temporal memory or phase information, agents cannot track or predict instantaneous flow states. Instead, each action-outcome pair provides a sample from the time-varying perturbation distribution. Through repeated interactions across training episodes, the agent effectively learns a policy robust to the statistical ensemble of flow states rather than responding to specific instantaneous configurations. This temporal integration emerges naturally from the physical averaging inherent in embodied interactions—the agent's learned behavior implicitly encodes the time-averaged flow statistics without requiring explicit memory mechanisms (detailed analysis in Supplementary Materials 7).

These results establish embodied intelligence as a robust navigation principle across diverse perturbation regimes. Whether facing steady vortices, reversed flows, or dynamic perturbations, successful policies emerge from the interplay between physical dynamics and learned adaptations. The agent effectively transforms its body-environment interactions into a computational resource, extracting and responding to hidden environmental information through the correlations embedded in its own motion dynamics.

\section{Discussion}


The experimental study with reinforcement learning demonstrates how proprioceptive cues enable an active microswimmer agent to counteract strong unobserved flow-perturbations, realizing microscale navigational autonomy in challenging conditions. This establishes a paradigmatic shift towards real-world realization of learning from body-environment interactions -- embodied dynamics, rather than solely relying on sensor-based information. This enables an adaptation process resulting from a feedback mechanism, where reinterpretation of the action outcomes in the environment encode flow-perturbations.

Each action inherently carries the imprint of the surrounding flow, as the net velocity already represents the combined effect of self-propulsion and perturbation. In absence of explicit flow sensing, proprioceptive cues encode these repeated hydrodynamic interactions through time-averaged correlations between its states, chosen actions, and the resulting transition probabilities. In stationary flow, the predictable correlation allows the agent to form compensatory policies. Furthermore, the emergence of mirror-image strategies for opposite flow directions confirms that embodied dynamics encode sufficient information for environmental discrimination -- an experimental validation of morphological computation principles \cite{hauser2011morphological,nakajima2015information}. This also highlights the intrinsic robustness of this learning paradigm: policies emerge from statistical structure of perturbations enabling generalization across different flow configurations, including dynamically varying fields.

Three technical insights emerge from our experimental realization. First, online learning converges within approximately 50 episodes, aligning with feasible experimental training timescales. Second, the use of five discrete action directions strikes a practical balance between control granularity and computational efficiency—crucial for resource-constrained, real-time operations at the microscale. Third, asymmetric rewards (stronger penalties for moving away) accelerate learning by discouraging excessive exploration, addressing a persistent challenge in physical RL systems. Collectively, these design choices illustrate how RL can be translated from abstract algorithms in simulations into real-time embodied controllers. Unlike previous RL-controlled microswimmers in controlled settings \cite{MuinosLandin2021,Bechinger2024_MARL}, our agent learns under genuinely hidden perturbations, using only position and displacement information - embodied interactions, effectively demonstrating that elaborate sensory systems are not prerequisites for microscale autonomy.

While this study demonstrates a proof of concept of embodied learning for microscale navigation, several limitations outline clear directions for advancement. Operating in quasi-2D environments with relatively simple flows, the current study demonstrates feasibility rather than full capability. Extending to three-dimensional turbulent flows with obstacles would better reflect real-world applications. In addition, action discretization and a single-agent setup restricts performance, while rapid flow variations may demand memory or recurrent architectures beyond the current averaging approach.

The path forward spans both fundamental and applied directions. While exploring different microswimmer platforms  - acoustic, magnetic or chemical allow learning from unique embodied dynamics, multi-agent scenarios may reveal collective embodied intelligence - advancing principles of swarm robotics~\cite{dauchot2023morphological}. Active reconfiguration of interaction through either modifying self-propulsion \cite{Rohde2025} or optimizing morphology through parametrizing interactions, offers further avenue for embedding computational capability and expand navigational versatility. Applied perspectives include targeted drug delivery, where such agents could learn to navigate complex physiological flows by exploiting transport dynamics rather than relying on pre-programmed responses~\cite{BioCargo_SciAdv_22}.

\section{Conclusion}

In summary, this study establishes that physical embodiment can serve as a computational resource for navigation of active microswimmers in challenging microfluidic environments, advancing the field of synthetic microrobots. Leveraging only on the proprioceptive cues and body-environment interactions, it is possible to encode sufficient information for adaptive navigation in presence of complex stationary or dynamic flows without explicit sensing. The ability to adapt to static and dynamic flow perturbations up to several times their propulsion speed highlights the potential of embodied intelligence for microscale navigation.

Our approach represents a significant step towards exploitation of morphological control rather than miniaturizing conventional sensors. The study advances the field from externally controlled or pre-programmed microrobots toward autonomous, physically informed intelligence, where morphology and hydrodynamic coupling play an integral role in information processing. Furthermore, this opens up design pathways for microswimmers that can leverage their shape, actuation modes, and interaction with surrounding flows for improved adaptability in real-world applications. This enables autonomous microsystems for environments where traditional sensing fails—from biomedical interventions to environmental monitoring—while bridging biological inspiration with practical implementation.



\section{Materials and methods}

\noindent \textbf{Sample:} The microswimmers used in the RL experiments consisted of gold nanoparticle coated melamine formaldehyde microswimmer (AuMF) of radius $R = \SI{1.09}{\micron}$ (microParticles GmbH). The gold nanoparticles (sizes vary between $8$-$\SI{30}{\nm}$) covers approximately $30\%$ surface area of the particles and facilitates the heating upon partial absorption of incident optical field. The encasing microscope glass coverslips were cleaned and spin-coated with $2\%$ (w/v) polystyrene solution and followed by a dipping in a $2\%$ (w/v) Pluronic F127 solution and then rinsed and dried. The thermally vapour deposited 50 nm Au film were only treated with pluronic followed by rinsed and dried for use. This prevents the micro particles from sticking to the substrate. A dilute aqueous suspension of the microswimmers and $\SI{3}{\micron}$ $\mathrm{SiO_2}$ spacer solution was prepared and $\SI{0.6}{\vol}$ was pippeted between the clean substrates. The substrate edges are sealed with PDMS to avoid leaking and evaporation.

\noindent\textbf{Setup:} The experimental setup used for the study consisted of a custom built inverted dark-field optical microscope (Olympus IX71) with a with a Piezo stage (Physik Instruments) placed on a stepper motor for coarse control. A LED (Thorlabs Solis 3C) module along with a dark-field oil-immersion condenser lens (Olympus, NA 1.2) was used as the illumination unit. The scattered light was collected using an oil-immersion objective lens ($100\times$, $0.6-1.3$ NA) with 0.8 NA and was projected onto an EMCCD (Andor iXon) using external optics. A $\SI{532}{\nm}$ laser was focused onto the sample using the imaging objective lens to in order to heat the active particles and it’s steering was achieved using an acousto-optic deflector (AOD; AA  Opto-Electronic) along with a 4-f system. The data communication between the AOD and the custom LabVIEW program was performed via an ADwin board (ADwin-Gold II, Jäger  Messtechnik). The field of view chosen for the realtime detection, analysis of particle position was $1000\times 1000$ pixel of the camera, translating into a spatial dimension of $\SI{75}{\micron}\times\SI{75}{\micron}$ with an average exposure time $\tau \approx \SI{160}{\ms}$ between consecutive actions (see Supplementary Materials 1).

\paragraph{Acknowledgement}
N. M. and N.S. are supported by BMBF (Federal Ministry of Education and Research) through
ACONITE (01IS22065). N.M., N.S. and FC are supported by the Center for Scalable Data Analytics and Artificial Intelligence
(ScaDS.AI.) Leipzig and by the European Union and the Free State of Saxony through BIOWIN.
N.M. is also supported by the Max Planck IMPRS CoNI Doctoral Program.

\bibliographystyle{unsrt}
\bibliography{references.bib}

\end{document}